\documentclass[
    ,final            
  ]
  {aipproc}

\layoutstyle{6x9}

\begin{document}

\title{Exciting Developments in Hadron Spectroscopy}

\classification{13.66.Bc,13.20.-V,13.25.-K,13.40.Gp}
\keywords      {Hadron Spectroscopy}

\author{Kamal K. Seth}{
  address={Department of Physics and Astronomy, 
        Northwestern University, Evanston, IL, 60208, USA}
}

\begin{abstract}
There has been a renaissance in hadron spectroscopy during the last couple of years.  Long lost states have been tracked down.  Unexpected states are showing up all over, and numerous measurements with unprecedented precision are being reported.  A review is presented.
\end{abstract}

\maketitle


\section{1. Introduction}

Heavy quark ($c,~b$) spectroscopy is the best way of studying Quantum Chromodynamics, QCD.  The $|c\bar{c}>$ charmonium and $|b\bar{b}>$ bottomonium states have several advantages over light quark ($u,~d,~s$) hadrons.  The relativistic effects are much smaller, the strong coupling constant $\alpha_S\approx0.2-0.3$ is small enough to permit use of perturbative methods to a large extent, and the spectra consist of clearly separated narrow states.  This is illustrated in Fig. 1 for charmonium.

Over the last three years there has been a renaissance in heavy--quark spectroscopy with genuine discoveries, important precision measurements, and a host of surprises.  These are the subject of this talk, but I also refer you to other related talks presented elsewhere in these proceedings \cite{othertalks}.

\begin{figure}[!tb]
\includegraphics[width=4.in]{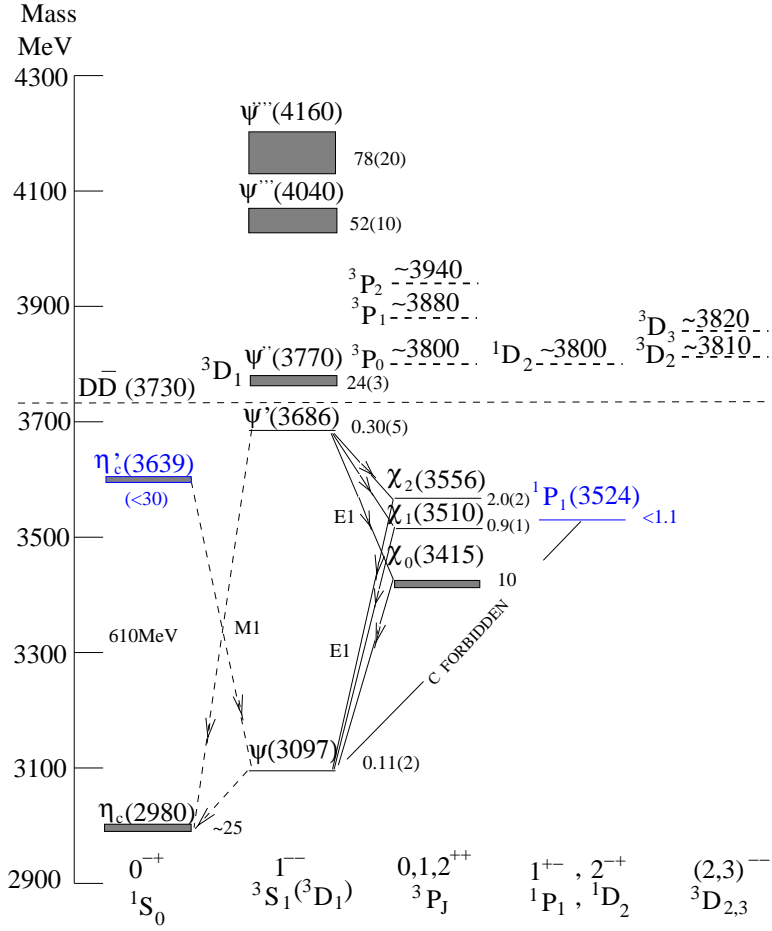}
\caption{Spectra of the states of Charmonium.}
\end{figure}

\section{2. Discovery of the Long--Lost Singlet States of Charmonium}

As is well known, the central $q\bar{q}$ interaction is well represented by the so--called Cornell potential (or its variants) as the sum of a one--gluon exchange Coulombic potential proportional to $1/r$ and a not--well--understood confinement potential, generally taken as Lorentz scalar and proportional to $r$ (Fig. 2).  Far less is known about the spin dependence of the $q\bar{q}$ potential which is responsible for the spin--orbit and spin--spin splittings of the $q\bar{q}$ states.  The crucial determinant of the spin dependent $q\bar{q}$ interaction is the hyperfine spin--singlet/spin--triplet splitting of the states.  Unfortunately, despite numerous and valiant efforts, no spin--singlet states have ever been successfully identified in bottomonium, and we have to depend exclusively on charmonium.  In charmonium the singlet $S$--state $\eta_c(1^1S_0)$ and the triplet $S$--state $J/\psi(1^3S_1)$ have been known for a long time, with the hyperfine splitting, $M(J/\psi)-M(\eta_c)=117\pm1$ MeV.  The other two bound singlet states $\eta_c'$, or $\eta_c(2^1S_0)$, and $h_c(^1P_1)$ have long eluded successful identification.  Well, now they have been finally discovered.
\vspace{12pt}

\begin{figure}[!tb]
\includegraphics[width=3.in]{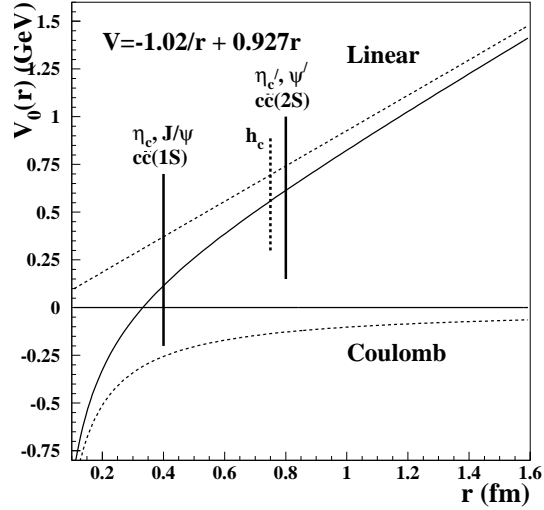}
\caption{Schematic of the QCD $q\bar{q}$ potential (solid line), and its Coulombic and confinement parts (dotted lines) seperately.  The approximate locations of the $1S$, $2S$ and $^1P_1$ states are shown.}
\end{figure}

\noindent \textbf{2.1 The Radial Excitation of the Charmonium Ground State, $\eta_c'$, or $\eta_c(2^1S_0)$}

\noindent As shown in Fig. 2, the $S$-wave radial excitations, $\psi'$ and $\eta_c'$ sample the confinement potential almost exclusively.  As a result, they provide our only insight into how the hyperfine interaction varies in going from $1S$ states ($\eta_c$ and $J/\psi$), with an almost 50--50 exposure to Coulombic and confinement potentials, to $2S$ states ($\eta_c'$ and $\psi'$), which lie in the confinement potential territory.

The mass of $\psi'$ is extremely well known, $M(\psi')=3686.093\pm0.034$ MeV, but $M(\eta_c')$ was unknown.  None of the earlier attempts to identify $\eta_c'$, which included Crystal Ball \cite{cb-etacp}, Fermilab E760/E835 \cite{fnal-etacp}, DELPHI \cite{delphi-etacp}, L3 \cite{l3-etacp}, and CLEO \cite{cleo-etacp} were successful.  The breakthrough came from an unexpected source, Belle \cite{belle-etacp}.  Since then several experiments have confirmed the identification of $\eta_c'$, and their results are summarized in Table 1.  Fig. 3 shows the results of CLEO \cite{cleo-2gam-etacp} and BaBar \cite{babar-etacp} for the reaction $\gamma\gamma\to K_SK\pi$.  While it is clear that $\eta_c'$ has been successfully identified, its mass still has substantial uncertainty, and its width remains unmeasured.  Using the world average $M(\eta_c')=3628.3\pm2.1$ MeV, we get $\Delta M_{hf}(2S)=47.8\pm2.1$, which means $\Delta M(2S)/\Delta M(1S)=0.41\pm0.02$.  Although a potential model calculation can always be found with just about any result, it is a fact that most predictions were that $\Delta M(2S)/\Delta M(1S)\approx0.65$.  So, here we have a result for theory to digest.  On the experimental side we have still much work to do, the first being to find the width of $\eta_c'$.  Together with $\Delta M_{hf}(2S)$, it will shed light on whether we are observing effects related to the confinement interaction or configuration mixing.
\vspace{12pt}

\begin{figure}[!tb]
\includegraphics[width=3.0in]{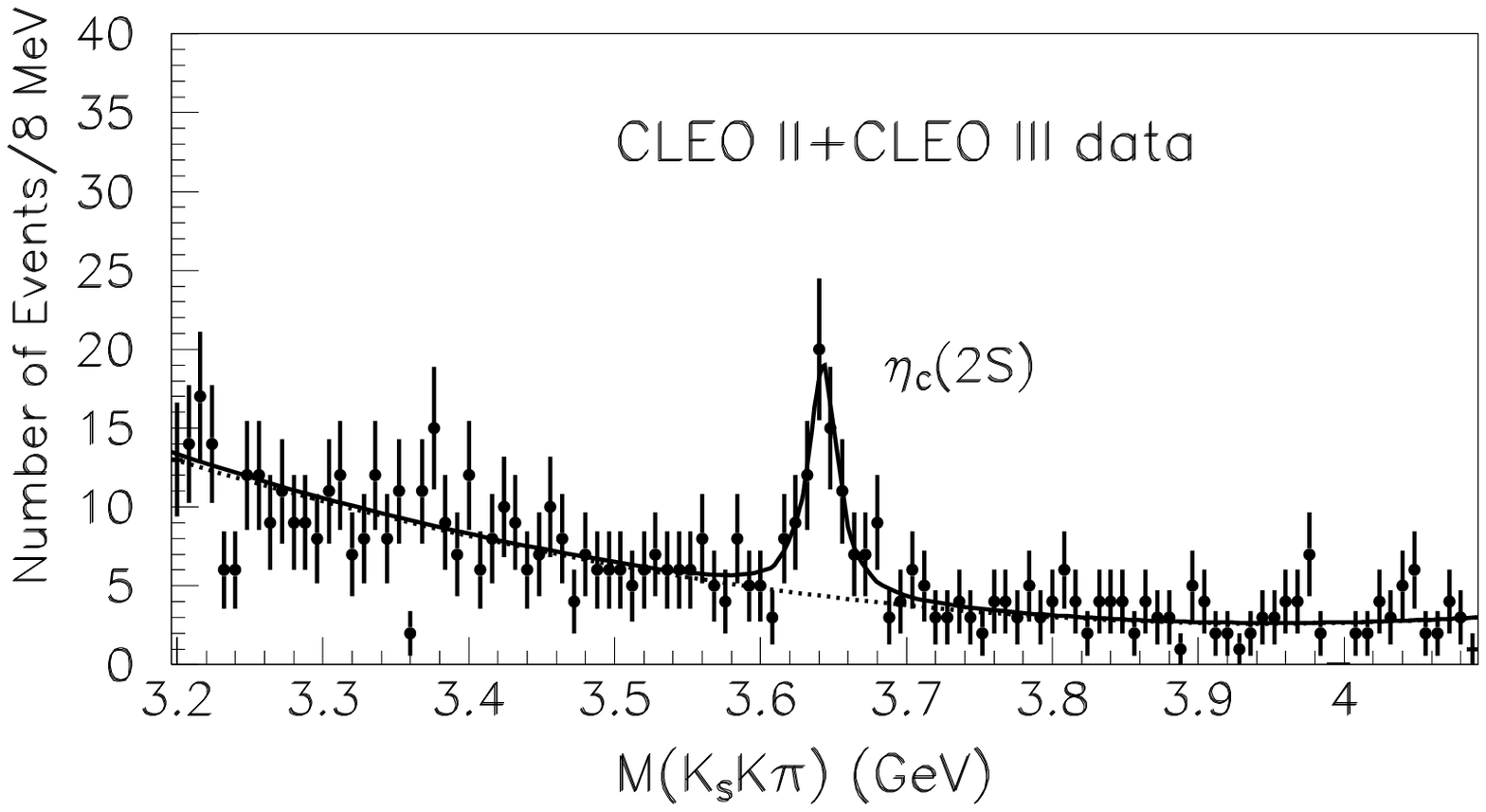}
\includegraphics[width=3.0in]{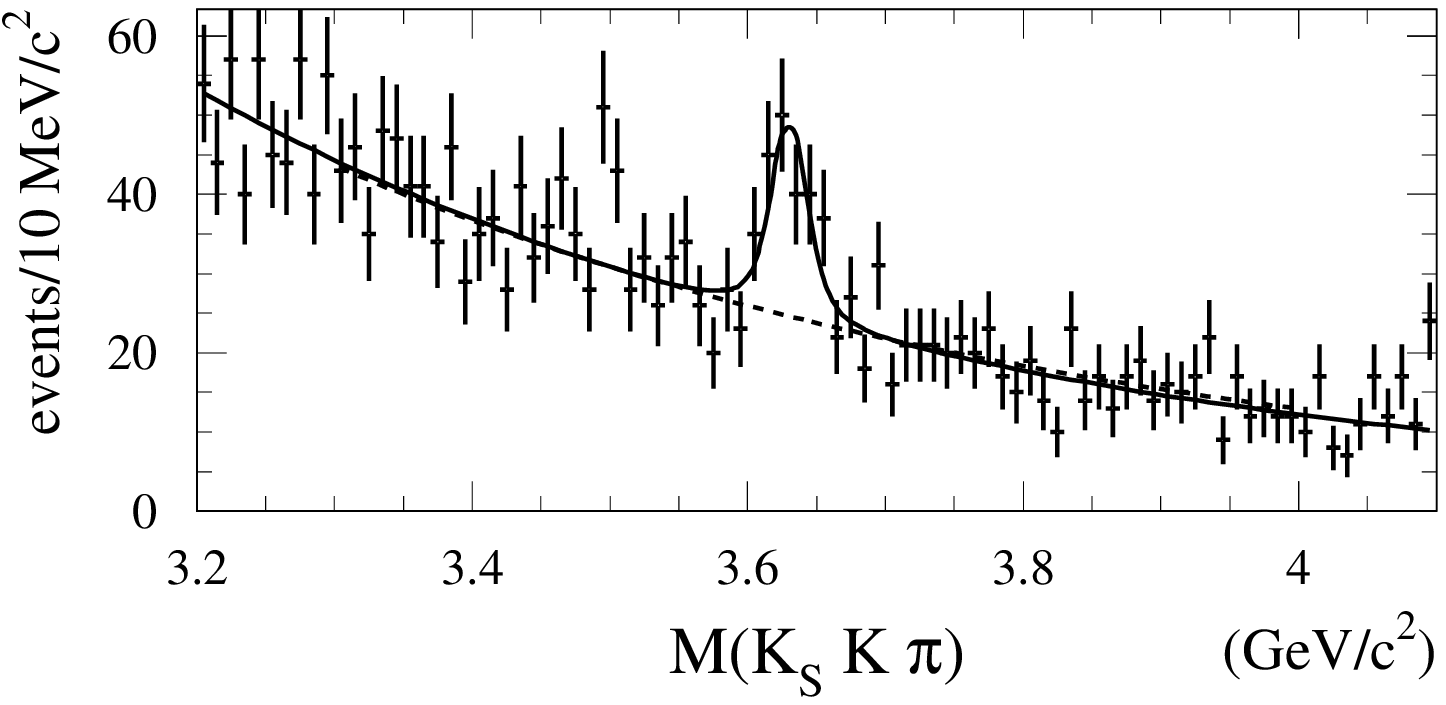}
\caption{Observation of $\eta_c'$ in the reaction $\gamma\gamma\to K_SK\pi$ by CLEO (left) and Babar (right).}
\end{figure}

\begin{table}[!tb]
\begin{tabular}{l|c|c|c}
\hline 
Measurement & N(cts) & M--2640 MeV & Width (MeV)\\
\hline
Belle \cite{belle-etacp} (2002): $B\to K(K_SK\pi)$ & $39\pm11$ & $14\pm10$ & $<55$ \\
Belle \cite{belle-etacp} (2005): $J/\psi + |c\bar{c}>$ & $311\pm42$ & $-14\pm9$ & $-$ \\
BaBar \cite{babar-etacp} (2004): $J/\psi + |c\bar{c}>$ & $121\pm27$ & $5\pm6$ & $22\pm14$ \\
BaBar \cite{babar-etacp} (2005): $\gamma\gamma\to K_SK\pi$ & $112\pm24$ & $-9\pm4$ & $17\pm9$ \\
CLEO  \cite{cleo-2gam-etacp} (2004): $\gamma\gamma\to K_SK\pi$ & $61\pm15$ & $+3\pm3$ & $6\pm14$ \\
\hline 
\end{tabular}
\caption{Summary of $\eta_c'$ observations.}  
\end{table}

\begin{figure}[!tb]
\includegraphics[width=2.4in]{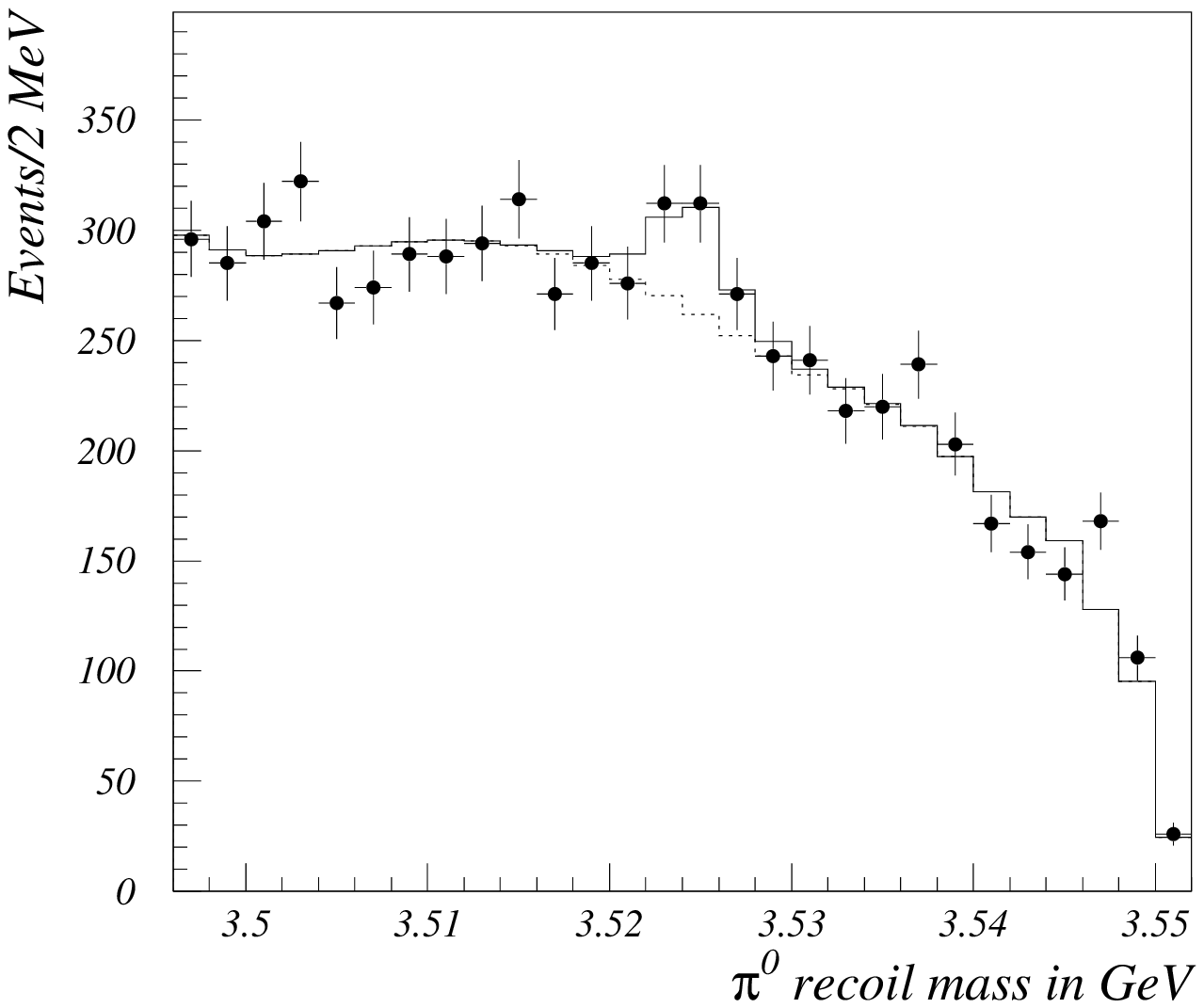}
\raisebox{1.8in}{\rotatebox{270}{\includegraphics[width=1.7in]{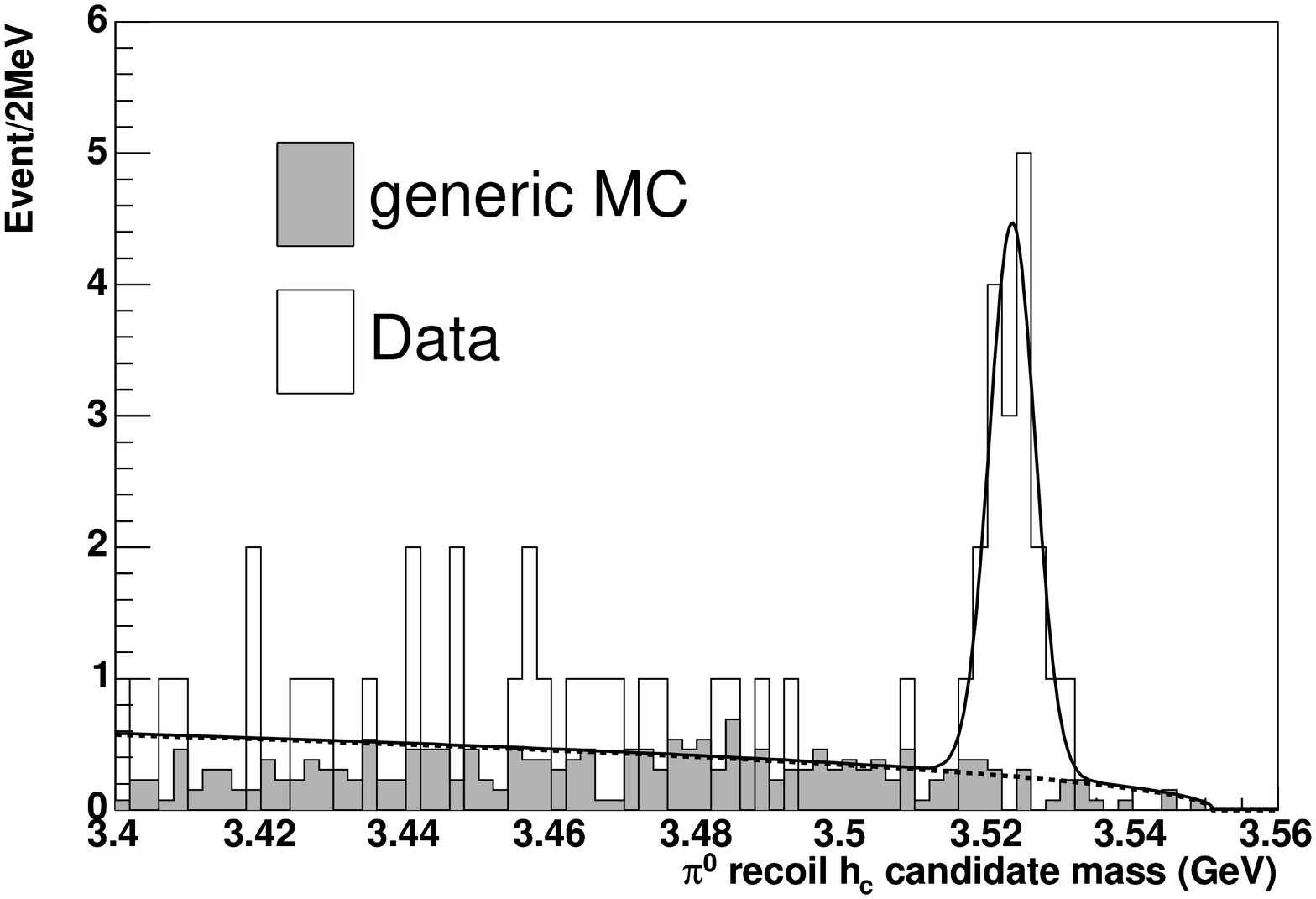}}}
\caption{Observation of $h_c(1^1P_1)$ in (left) inclusive analysis and (right) exclusive analysis at CLEO \cite{cleo-hc}.}
\end{figure}

\noindent \textbf{2.2 The Singlet P-wave State of Charmonium, $h_c(1^1P_1)$}

\noindent The singlet $P$--wave $h_c(1^1P_1)$ relates to a different aspect of the hyperfine interaction than $\eta_c(2^1S_0)$ does.  Here the question is whether or not there is a long range $\vec{s}_1\cdot\vec{s}_2$ interaction in the $q\bar{q}$ system.  In absence of such an interaction, the lowest order pQCD prediction is that the hyperfine, singlet--triplet splitting is finite only for S--wave states, and is identically zero for all higher $L$, i.e. $\Delta M_{hf}(L\ne0)=0$.  In particular, $\Delta M_{hf}(1P)=\left< M(^3P_J) \right>-M(^1P_1)=0$.  The centroid of the $^3P_J(J=0,1,2)$ states has been measured very accurately by the Fermilab E835 experiment, $\left<M(^3P_J)\right>=3525.36\pm0.06$ \cite{chimass}.  What is needed is a firm identification of $h_c(1^1P_1)$ and a precise measurement of its mass.  CLEO \cite{cleo-hc} has just reported the unambigous ($6\sigma$) observation of $h_c$ in a ``tour de force'' measurement of the reaction $\psi(2S)\to\pi^0h_c,~h_c\to\gamma\eta_c$.  Both inclusive and exclusive measurements of $M(h_c)$ in recoils against $\pi^0$ have been made.  In the inclusive measurement, either $E_\gamma$ or $M(\eta_c)$ are constrained.  In the exclusive measurement, $\eta_c$ is instead identified in seven hadronic decay channels.  Both inclusive and exclusive spectra are shown in Fig. 4.  The result is $\Delta M_{hf}(1P)=+1.0\pm0.6\pm0.4$ MeV (for more details, see my parallel session talk).  E835 \cite{e835-hc} has also just reported the results of the analysis of their 1996+2000 data for the reaction $\bar{p}p\to h_c \to \gamma\eta_c$.  They report $\Delta M_{hf}(1P)=-0.4\pm0.2\pm0.2$ MeV at a significance level of $\sim3\sigma$.

Two conclusions can be derived from these measurements.  The first is that the ``naive'' pQCD predication $\Delta M_{hf}(1P)=0$ is not being violated in any substantial manner, as had been feared by many theorists.  The second conclusion is that the all--important sign and the small magnitude of $\Delta M_{hf}(1P)$ is not yet pinned down.  In the near future, larger $\psi'$ running at CLEO is expected to reduce the errors in the CLEO measurement, and GSI, whenever it becomes operational, may be expected to do even better.

\section{3. A Sampling of Precision Charmonium Results}

Spectroscopy does not progress only by discoveries.  It makes some of its best gains by precision.  Recently, several such gains have been made.  At Novosibirsk \cite{novo-mass}, masses of $J/\psi$ and $\psi'$ have been measured at $\pm12$ keV and $\pm25$ keV levels respectively; a precision of $\sim 5$ parts per million.  At Fermilab \cite{chimass}, masses of the $\chi_J$ states have been measured with precision of 50--100 parts per million and widths to better than $\pm10\%$.  At CLEO \cite{cleo-psi-ll, cleo-psi-x}, lepton universality has been confirmed at a $\pm1\%$ level in $J/\psi$ decays to $e^+e^-$ and $\mu^+\mu^-$.  At CLEO \cite{cleo-psi-x}, isospin conservation has been confirmed at a $\pm2\%$ level in $\psi'$ decay to $\pi^+\pi^-J/\psi$ and $\pi^0\pi^0J/\psi$, and isospin violation has been observed at a $\pm0.4\%$ level in $\psi'$ decay to $\pi^0J/\psi$ and $\eta J/\psi$.  Also at CLEO \cite{cleo-psirad,cleo-chi2gam}, the long standing discrepancy between two--photon widths of the $\chi_2$ state determined by two--photon fusion and by two--photon decay has been resolved by new precision measurements of $\Gamma_{\gamma\gamma}(\chi_{c2})$ and $\mathcal{B}(\chi_{c2}\to\gamma J/\psi)$.

\begin{figure}[!tb]
\includegraphics[width=3.in]{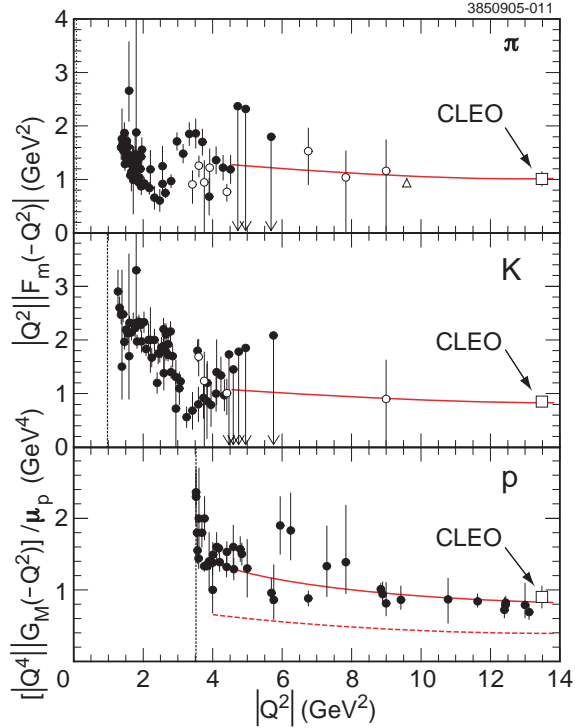}
\caption{CLEO results for the timelike form factors at $Q^2=13.48$ GeV$^2$ of the pion (top), kaon (middle), and proton (bottom).}
\end{figure}

\section{4. Timelike Form Factors of Pion, Kaon, and Proton}

Electromagnetic form factors of hadrons provide deep insight into their quark structure and help define the domain of validity of pQCD.  Except for the magnetic form factor of the proton for spacelike momentum transfers, few measurements of spacelike or timelike form factors of any hadrons exist at large enough momentum transfers to shed light on the highly controversial debate on the validity of pQCD at modestly large momentum transfers.  The Fermilab \cite{fnal-ff} discovery that the timelike form factor of the proton is nearly a factor two larger than the spacelike form factor of the proton in the entire region $Q^2=5-13$ GeV$^2$ tells us to expect surprises in the measurement of other form factors.  Unfortunately, the sparse data which exist for the form factors of pions and kaons (which often could not be separately identified) are essentially limited to $Q^2\le4$ GeV, and have very large (up to 100\%) errors.  This situation has been remedied recently by a very demanding measurement made at CLEO \cite{cleo-ff} for $Q^2=13.48$ GeV$^2$.  To get an idea of how difficult these measurements are, it is enough to point out that at this momentum, $e^+e^-$ collisions produce about 500 times more muon pairs than pion pairs, and one must distinguish between them.  The results of this CLEO measurement with less than $\pm10\%$ errors are shown in Fig. 5.

\section{5. The Suprising States}

Let me move on to the unexpected states which seem to be popping up all over during the last two years.  The veterans among these are X(3872) and the Pentaquark.  The newcomers are X(3943), Y(3943), Z(3931), and V(4260).  This proliferation is exciting, but also rather baffling.  It arises primarily from the fact that huge integrated $e^+e^-$ luminosities ($\ge300$ fb$^{-1}$) are now available at Belle and BaBar, and very weak resonances are showing up.  It will be a while before the dust settles down and we really know what is going on.  Let me describe these surprises, and their current status, one by one.
\vspace{12pt}

\noindent \textbf{5.1 The Pentaquark}

\noindent As is probably well known to most of the audience, when the Pentaquark was born in 2003, it caused great interest.  Google tells me that there are 31,300 entries for it by now.  There were many reported sightings of pentaquarks of all kinds, and even a greater number of reported failures to find the expected signals. Finally, there is the recent JLab report of the absence of the pentaquark signal in a large statistics repeat of their earlier measurement.  This reminds me of a similar history of claimed observations of dozens of dibaryons, which all eventually evaporated \cite{kamdibaryon}. My personal, perhaps biased, conclusion is that the pentaquark is now on life--support.  May it rest in peace!  See the talk by E. Smith \cite{esmith} for more information.
\vspace{12pt}

\noindent \textbf{5.2 The X(3872) Mystery}

\noindent In 2003 Belle \cite{bellex} announced the discovery of an unexpected state, X(3872).  It was quickly confirmed by CDF \cite{cdfx}, D\O~\cite{d0x} and BaBar \cite{babarx} (see Fig. 6).  The decay X(3872)$\to\pi^+\pi^-J/\psi$ is the dominant decay.  The average of the masses measured by the four experiments is $M(\mathrm{X})=3871.5\pm0.4$ MeV.  Note that $M(D^0)+M(\overline{D^{*0}})=3870.3\pm2.0$ MeV \cite{pdg}.  The best measurement of the width gives $\Gamma(\mathrm{X})\le2.3$ MeV.

The unique decay, the narrow width, and the closeness of its mass to $M(D^0D^{*0})$ have given rise to intense theoretical speculations about the nature of X(3872).  Theoretical speculations are that X(3872) is a charmonium state ($1^{++}$, $2^{--}$, $3^{--}$), a hybrid ($1^{++}$), a glueball mixed with vector charmonium ($1^{--}$), a $D^0D^{*0}$ ($1^{++}$, $0^{-+}$) molecule?  Perhaps the most provocative of these proposals is the molecular model, because no $|q\bar{q}>\cdot|q\bar{q}>$ molecules have ever been found! To sift through these speculations it is necessary to determine $J^{PC}$(X). Frantic searches are in progress at Belle and BaBar for decays to establish $J^{PC}$(X). 

\begin{figure}[!tb]

\setlength{\baselineskip}{0pt}
\begin{tabular}{cc}
\includegraphics[width=2.0in]{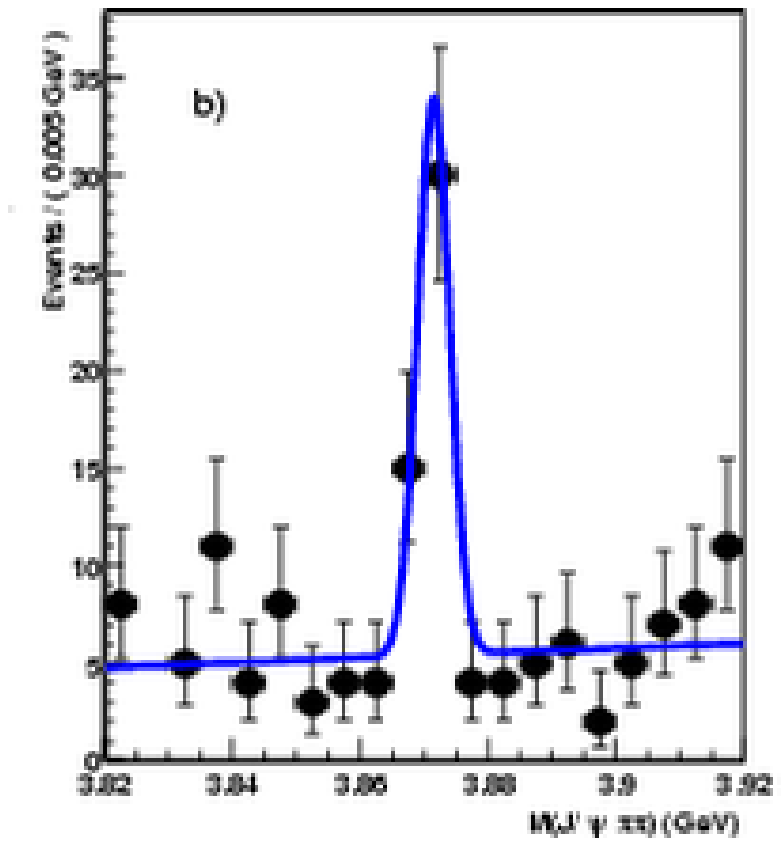}
&
\includegraphics[width=2.2in]{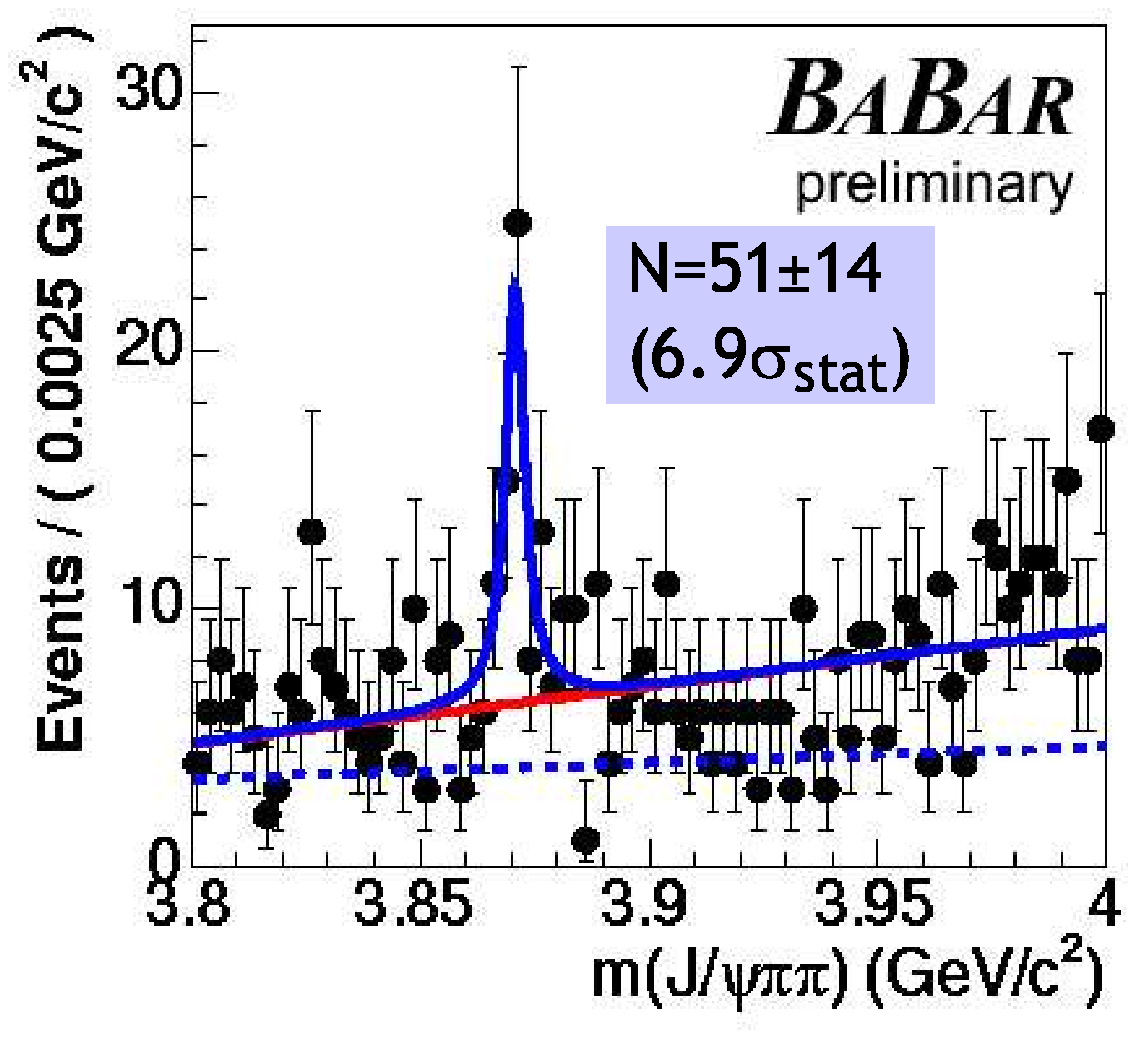}
\\
(a) & (b) \\
\includegraphics[width=2.0in]{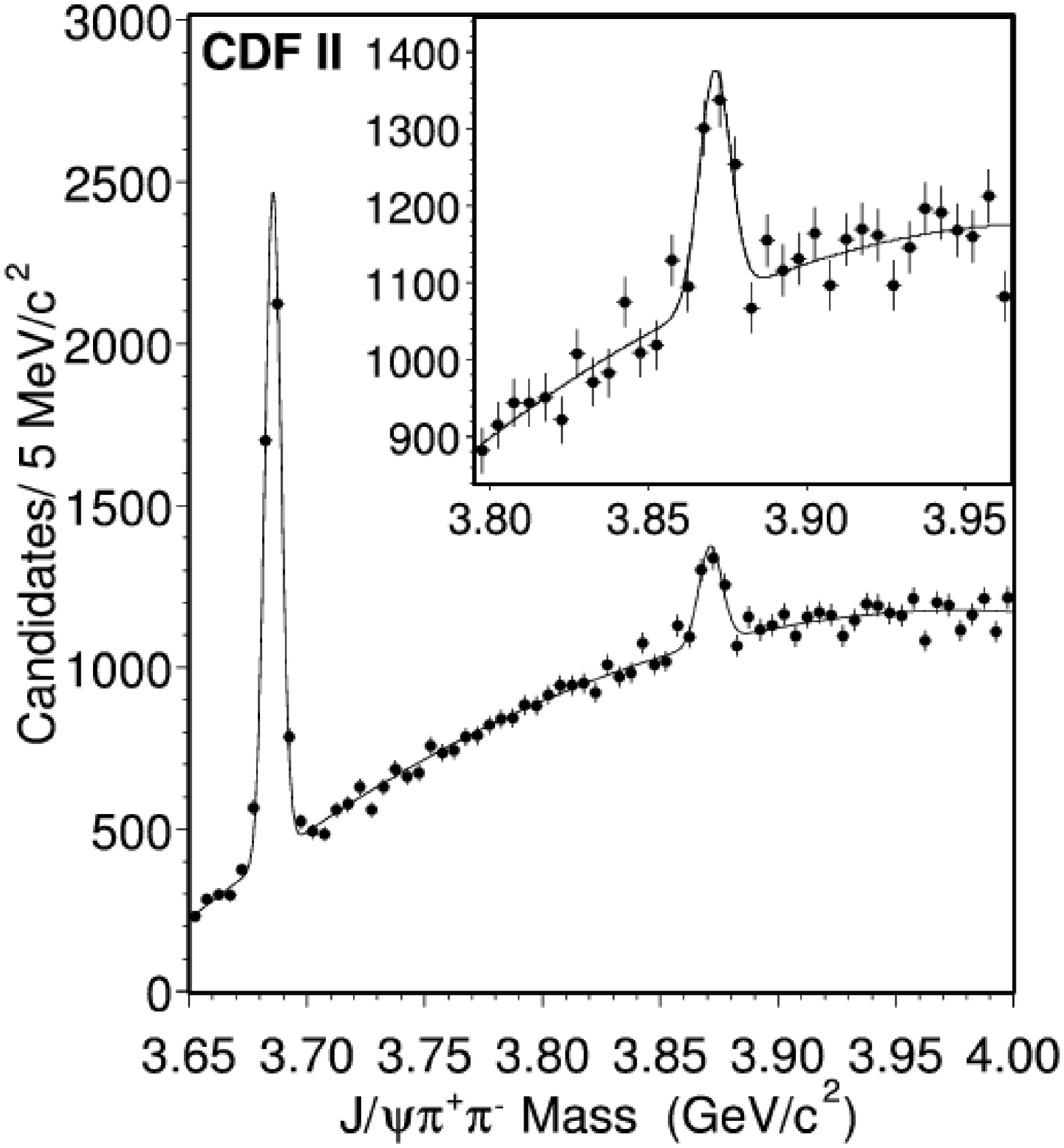}
&
\includegraphics[width=2.2in]{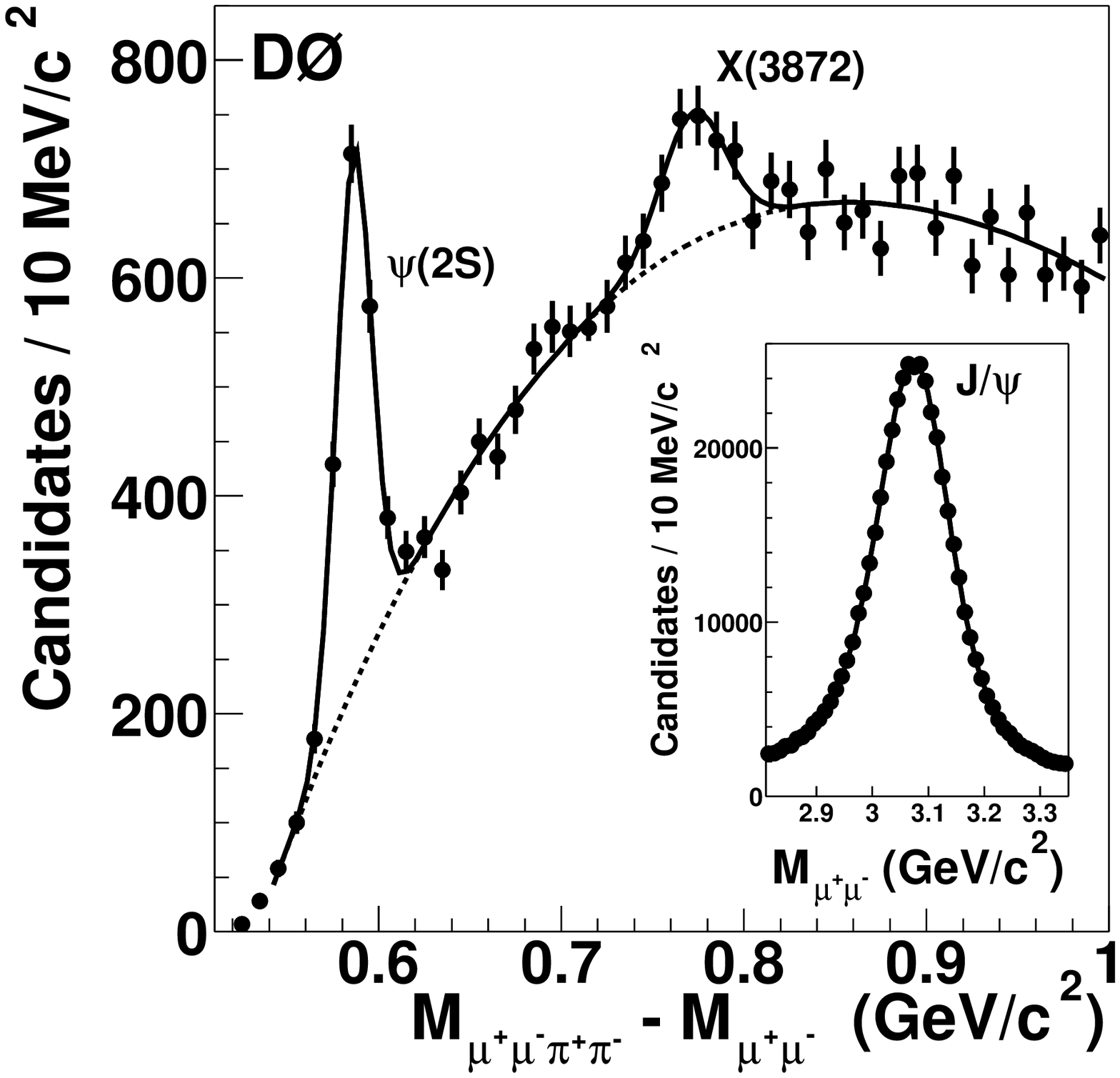}
\\
(c) & (d) \\
\end{tabular}

\caption{Observations of X(3872) by (a) Belle \cite{bellex}, (b) CDF \cite{cdfx}, (c)D\O~\cite{d0x}, and (d) BaBar \cite{babarx}.}
\end{figure}

\begin{figure}[!tb]
\includegraphics[width=2.5in]{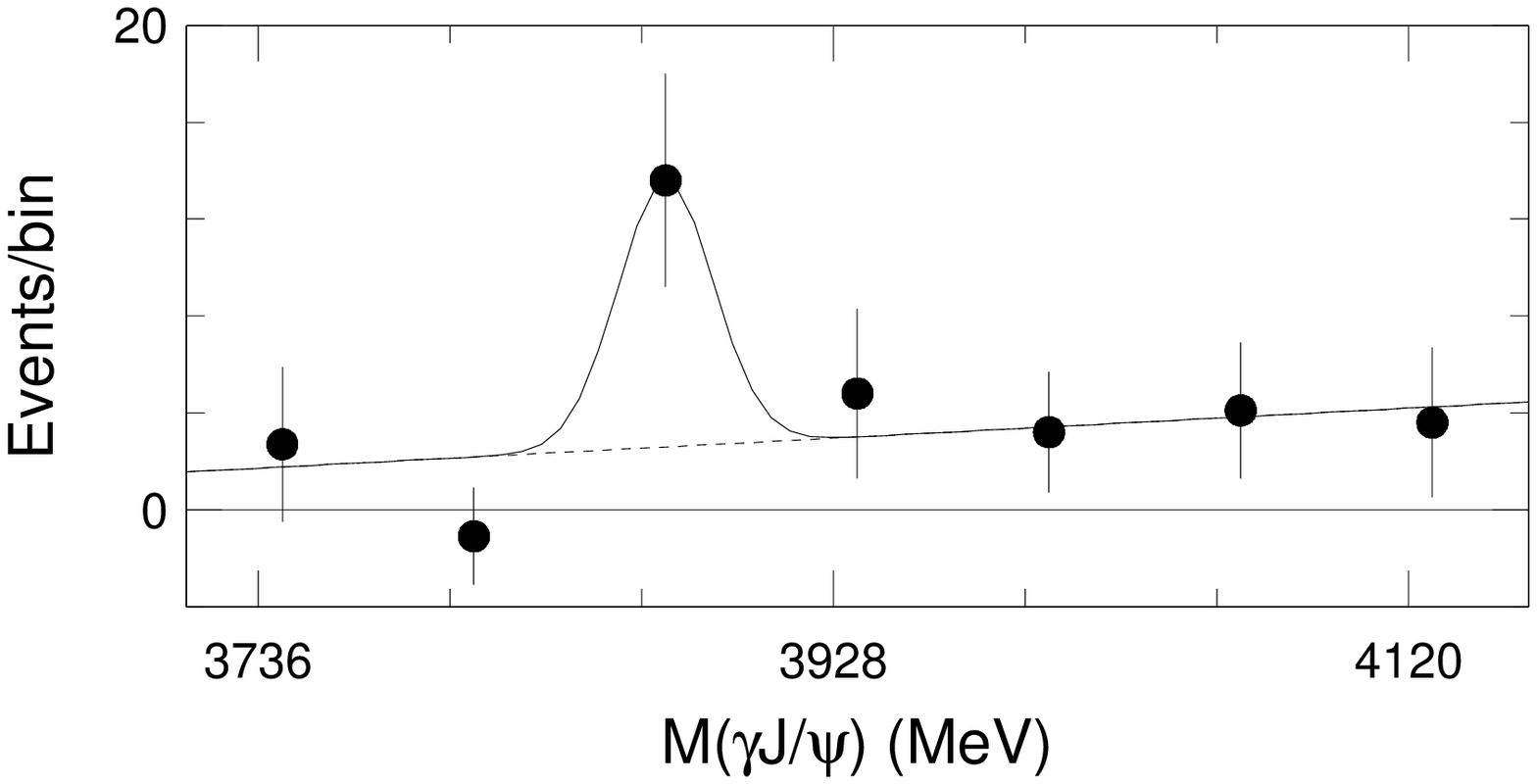}
\includegraphics[width=2.4in]{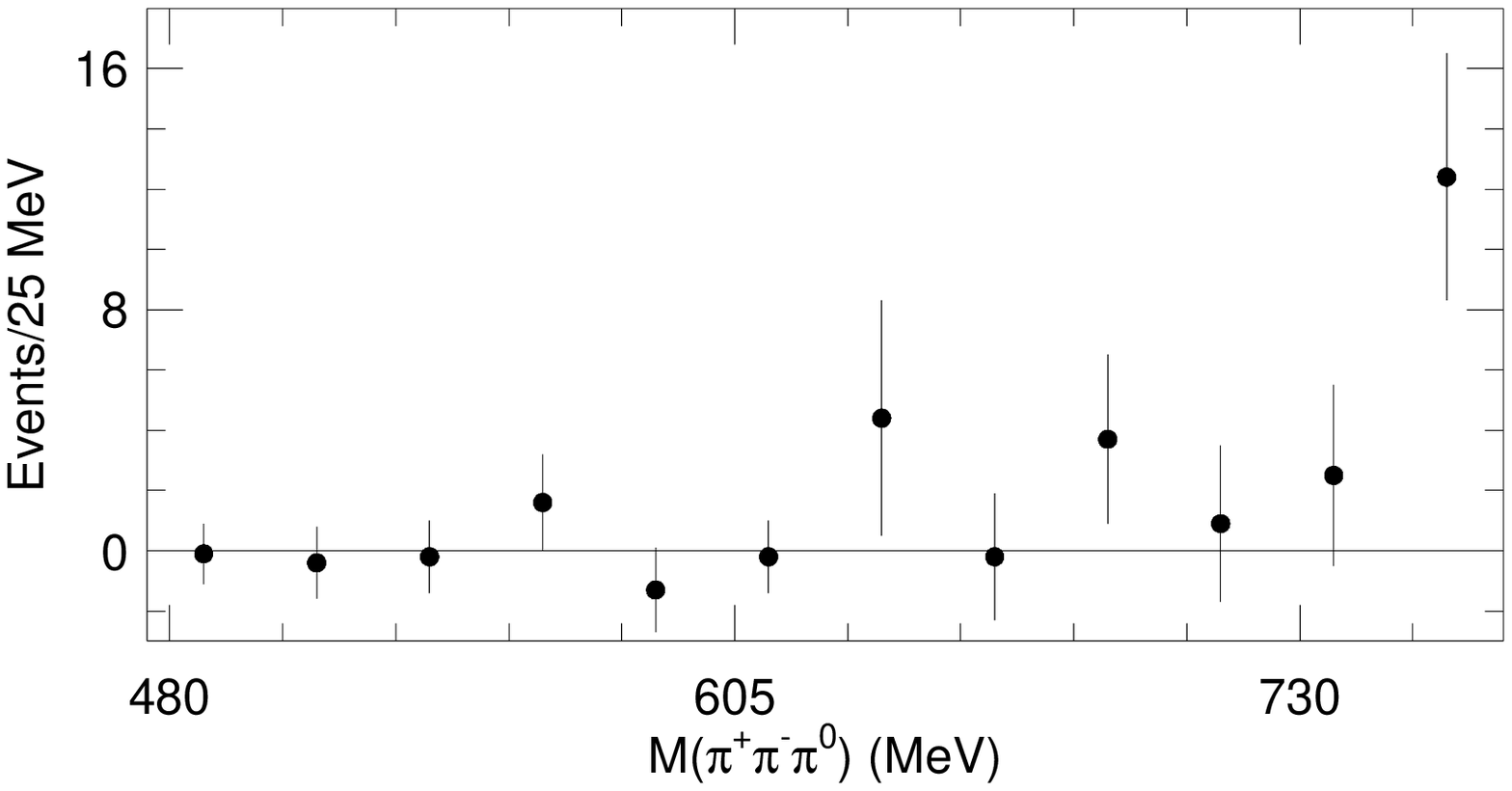}
\caption{Observations by Belle \cite{bellex2} of  $\mathrm{X}\to\gamma J/\psi$ (left) and $X\to \omega J/\psi$ (right).}
\end{figure}

This state continues to defy understanding.  CLEO \cite{cleox}, BES \cite{besx}, and Babar \cite{babarisr} have put the limit, $\mathcal{B}(\mathrm{X}\to\pi^+\pi^- J/\psi)\Gamma(\mathrm{X}\to e^+e^-)<6$ eV, which, depending on how large $\mathcal{B}(\mathrm{X}\to\pi^+\pi^-J/\psi)$ is, might weigh in against X being a vector.  However, D\O~finds $\mathrm{X}\to\pi^+\pi^-J/\psi$ decays to have all the same characteristics as the vector $\psi(2S)$ decays to $\pi^+\pi^-J/\psi$.  Belle \cite{bellex2} has presented arguments against the charge parity of X being negative on the basis of observations of $\mathrm{X}\to\gamma J/\psi$ and $X\to \omega J/\psi$.  However, as shown in Fig. 7, the two observations consist of $13.6\pm4.4$ and $12.1\pm4.1$ counts (each in one mass bin), respectively.  CDF \cite{cdfx} reports the $\pi^+\pi^-$ mass distribution to be consistent with X being a $^3S_1$ vector with $C=-1$, or a $C=+1$ object decaying into $\rho J/\psi$.  Finally, potential model calculations do not really rule out $^3D_2(2^{--})$ and $^3D_3(3^{--})$ states at 3872 MeV,  In view of all this, I personally believe that all options are still open for X(3872).
\vspace{12pt}

\noindent \textbf{5.3 The Saga of X, Y, Z at $M\approx3940$ MeV}

\noindent There are three new states reported by Belle with masses which are statistically consistent with being identical \cite{belle-x, belle-y, belle-z}.  The spectra in which they were observed are shown in Fig. 8, and their characteristics are summarized in Table 2.  Each is formed in a different reaction and decays dominantly in a different channel.  X(3943) is observed in a recoil mass spectrum in which only $J=0^\pm,~C=+$ states $\eta_c(0^{-+}),~\eta_c'(0^{-+}),~\chi_{c0}(0^{++})$ are seen.   This would suggest $J(\mathrm{X(3943)})=0^\pm,~C=+$. The state Z(3931) with width similar to that of X(3943) is produced in $\gamma\gamma$ fusion, which guarentees $C=+$, but is found to have the $D(\bar{D})$ angular distribution characteristic of $J(\mathrm{Z(3931)})=2$.  The fitted width of Y(3942) is claimed to be $\sim4$ times larger than that of X or Z, and its decays are almost opposite to those of X(3943).  All this produces a very confusing picture because it appears to be rather implausible that three distinct states exist within a 10 MeV mass interval.  But Nature can always spring surprises!  In any case, we can look forward to additional data, particularly from BaBar, to sharpen the picture.
\vspace{12pt}

\begin{figure}[!tb]
\begin{tabular}{ccc}
\includegraphics[width=2.5in]{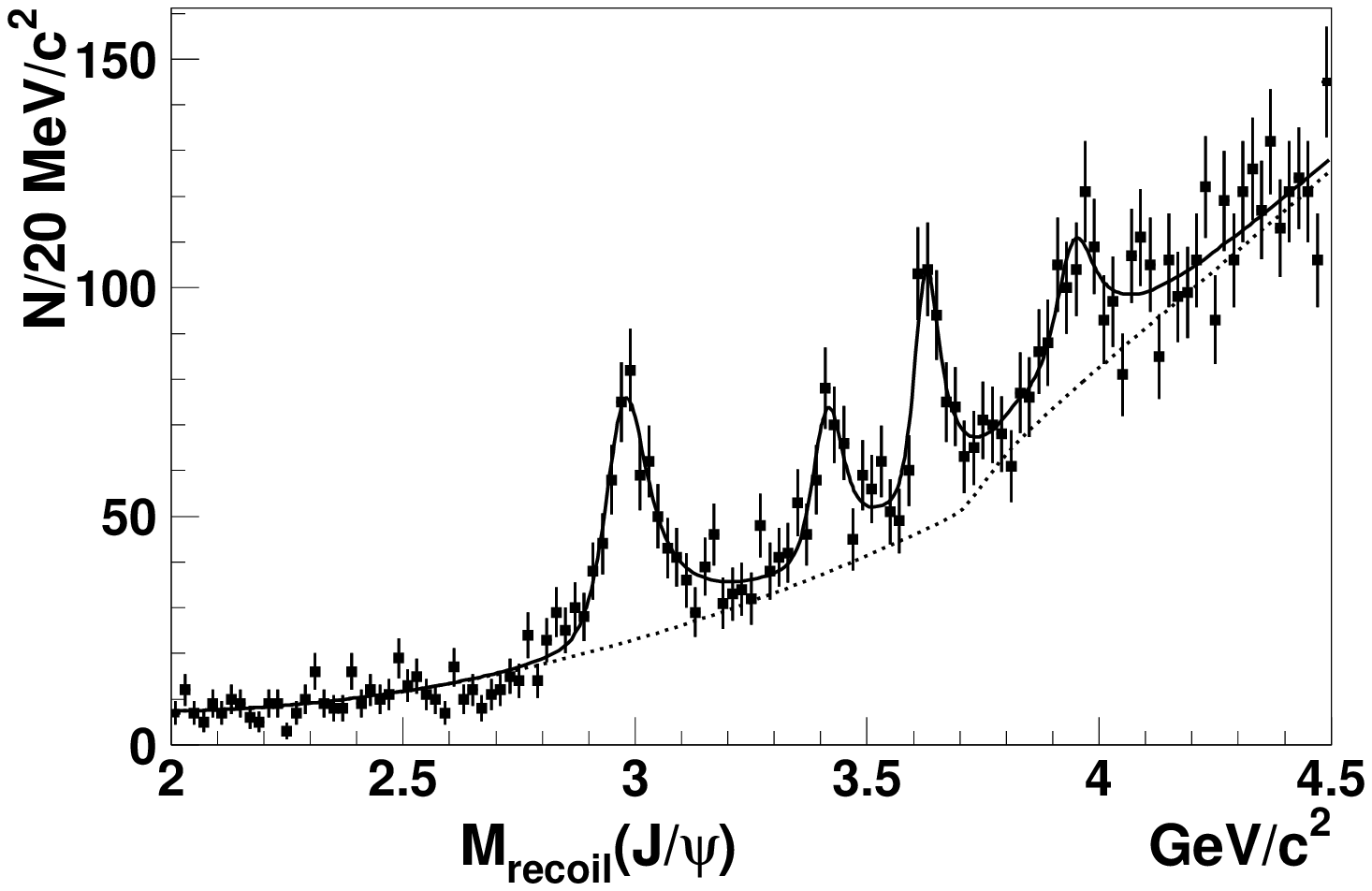}
&
\includegraphics[width=1.7in]{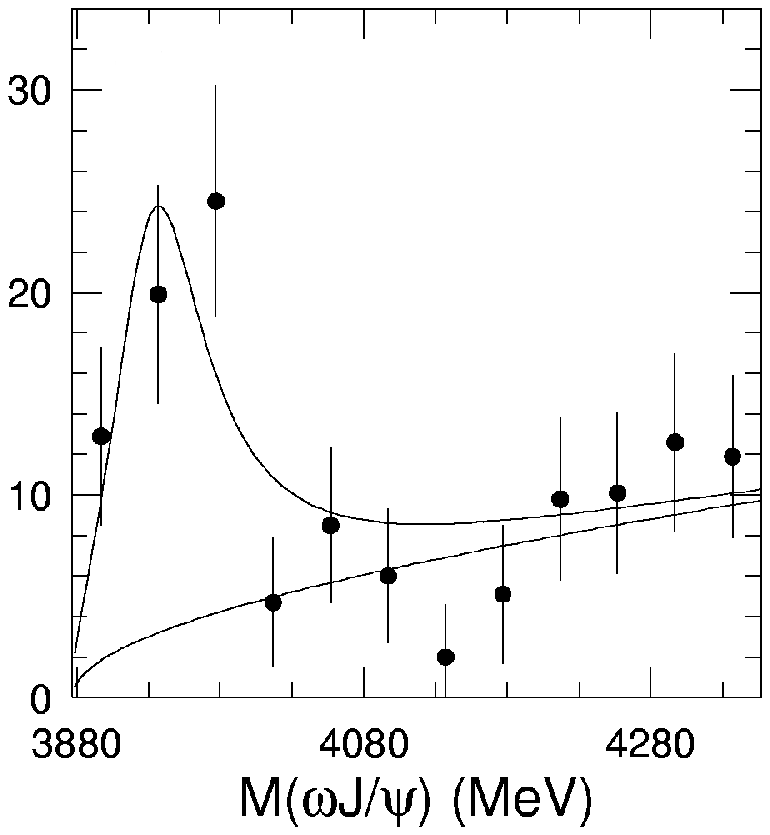}
&
\includegraphics[width=1.7in]{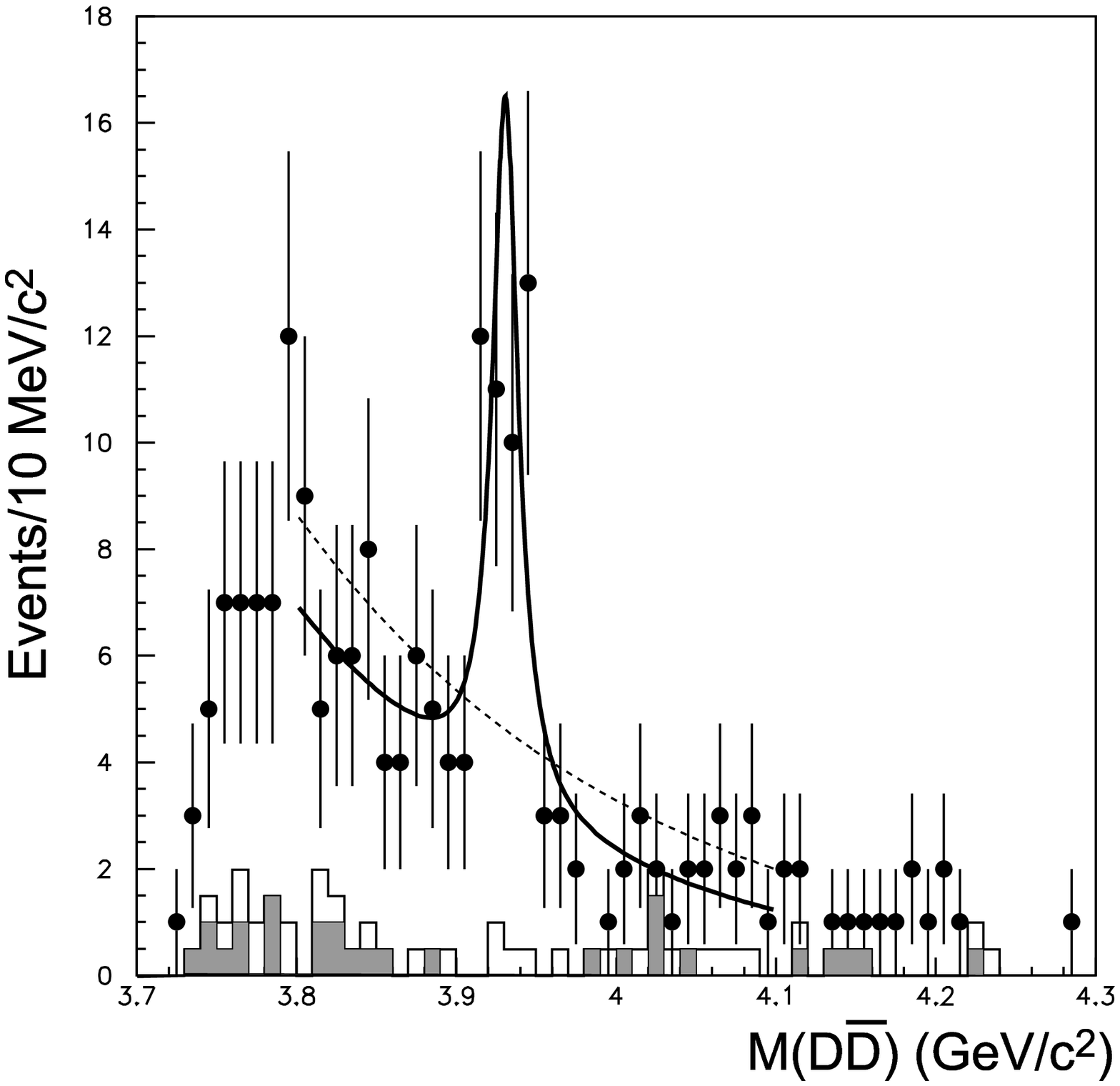}
\\
(a) X(3943) & (b) Y(3943) & (c) Z(3931) \\
\end{tabular}
\caption{Obervations of three new states by Belle.}
\end{figure}

\begin{table}[!tb]
\begin{tabular}{c|l|c|c|c|c|c}
\hline
 & \centering M(MeV) & $\Gamma$(MeV) & Formed in & Decays in & not in & suggests \\
\hline
X \cite{belle-x} & $3943\pm6\pm6$ & $15\pm10$ & $e^+e^-\to J/\psi(c\bar{c})$ & $D^*\overline{D}$ & $D\overline{D}$, $\omega J/\psi$ & ? \\
Y \cite{belle-y} & $3943\pm11\pm13$ & $87\pm22$ & $B\to K(\omega J/\psi)$ & $\omega J/\psi$ & $D^*\overline{D}$(?) & $c\bar{c}$ hybrid?  \\
Z \cite{belle-z} & $3931\pm4\pm2$ & $20\pm8\pm3$ & $\gamma\gamma$ fusion & $D\overline{D}$ & & $\chi_{c2}'(2^3P_2)$ \\
\hline
\end{tabular}
\caption{Summary of properties of three new states observed by Belle \cite{belle-x,belle-y,belle-z}.}
\end{table}
 
\begin{figure}[!tb]
\begin{tabular}{ll}
\includegraphics[width=3.1in]{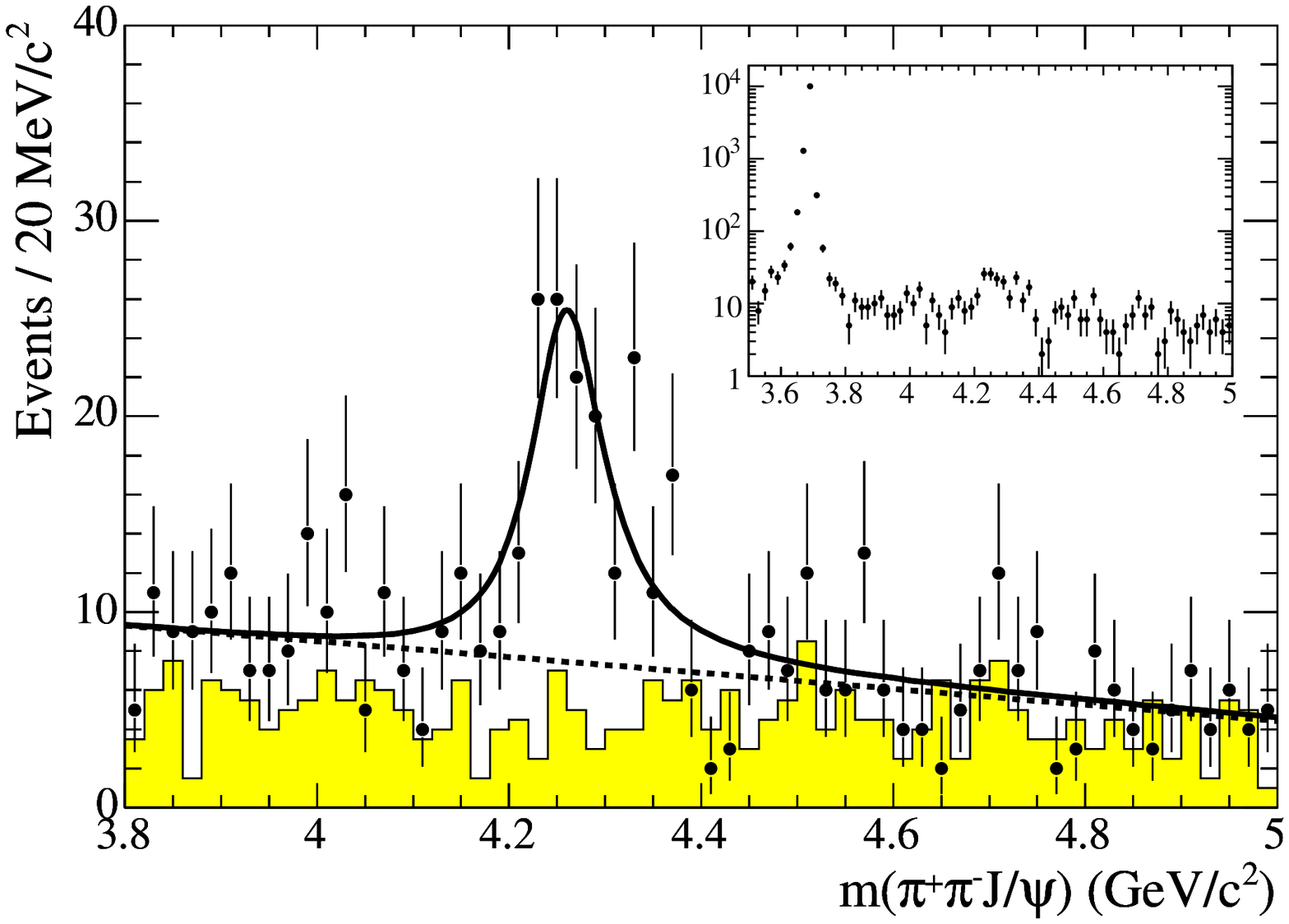}
&
\includegraphics[width=2.3in]{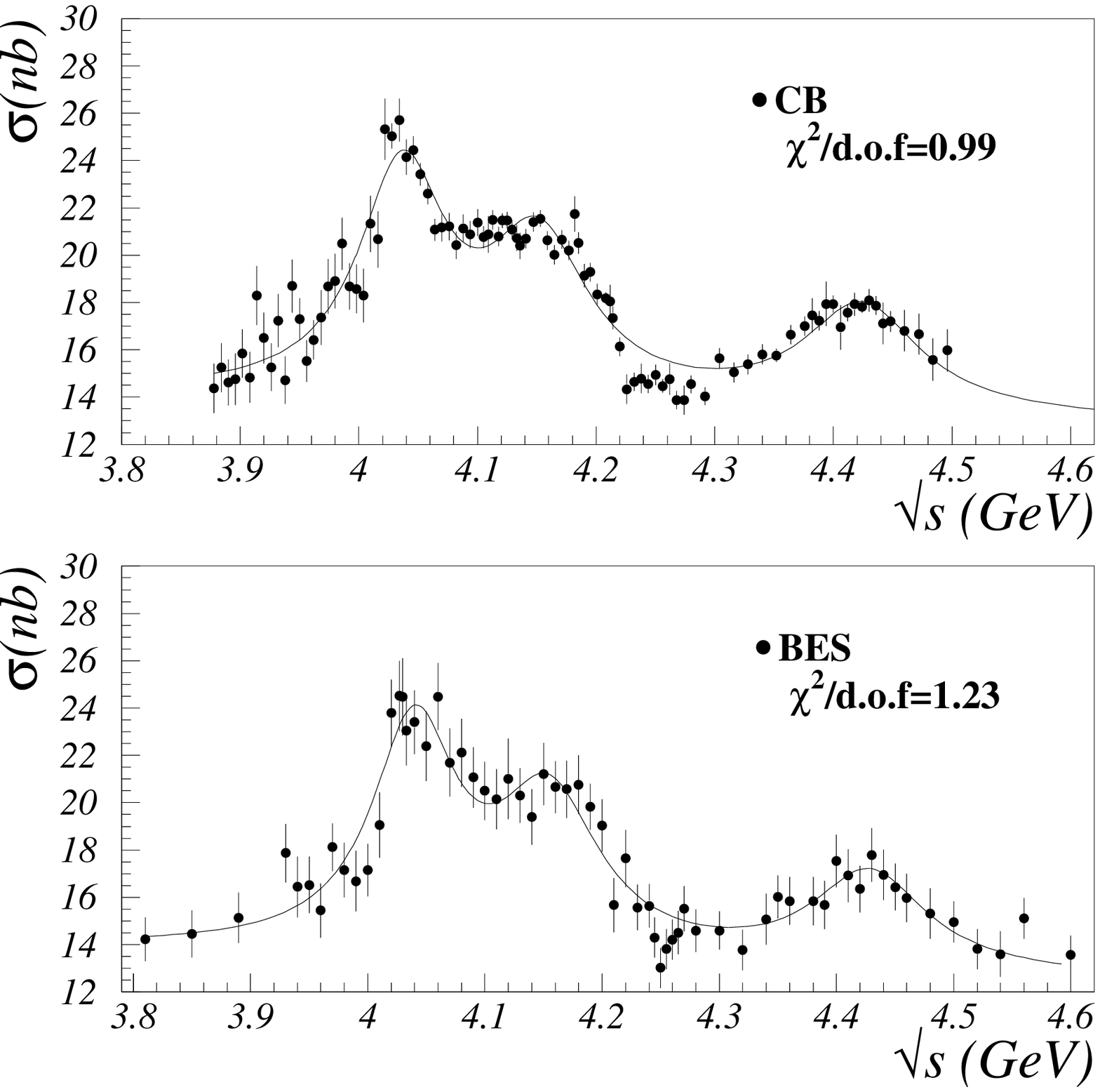}
\end{tabular}
\caption{(left) Observation by Babar \cite{babarisr} of V(4260) enhancement in ISR, (right) $\sigma(e^+e^-\to\mathrm{hadrons})$ \cite{rmeas} with minimum at $\sqrt{s}=4260$ MeV.}
\end{figure}


\begin{figure}[!b]
\includegraphics[width=3.7in]{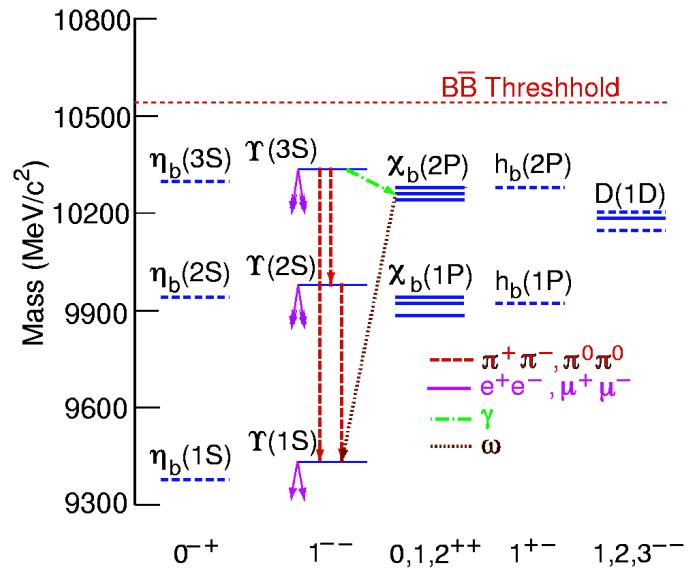}
\caption{Spectra of the bound states of Bottomonium.}
\end{figure}

\noindent \textbf{5.4 Three is Not Enough---Bring in the Fourth, V(4260)}

\noindent Belle has been running away with too many new resonances.  Now comes BaBar.  They have analyzed ISR events from 211 fb$^{-1}$ of data \cite{babarisr}, and report a convincing (significance $8\sigma$), broad enhancement in the invariant mass $M(\pi^+\pi^- J/\psi)$ spectrum (Fig. 9).  Since production via ISR guarentees a vector, and X, Y, and Z have been overused, I take the liberty of christening this state as V(4260).  The parameters of this enhancement are $M(\mathrm{V})=4259\pm8^{+2}_{-6}$ MeV, $\Gamma$(V)=$88\pm23^{+6}_{-4}$ MeV, $N=125\pm23$ events.  From this, BaBar measures $\Gamma(e^+e^-\to \mathrm{V})\times\mathcal{B}(\mathrm{V}\to \pi^+\pi^- J/\psi)=(5.5\pm1.0^{+0.8}_{-0.7})$ eV.  They suggest that it might be a previously unobserved $1^{--}$ resonance.  This is quite surprising because no vector around this mass is predicted, and the R measurements actually show a dip in this mass region as illustrated in the adjoining figure from Ref. \cite{rmeas}.

\section{6. Bottomonium}

The Bottomonium spectrum is shown in Fig. 10.

The world's largest sample of 21 million $\Upsilon(1S)$, 9 million $\Upsilon(2S)$, and 6 million $\Upsilon(3S)$ comes from CLEO.  These data sets have been analyzed to yield interesting new results and improved precision.
\begin{itemize}
\item For the first time, a non--$\pi\pi$ hadron transition between bottomonium resonaces has been observed \cite{omega-y1s}, with $\mathcal{B}(\chi_{b1}^\prime\to\omega\Upsilon(1S))=(1.63\pm0.38)\%$ and $\mathcal{B}(\chi_{b2}^\prime\to\omega\Upsilon(1S))=(1.10\pm0.34)\%$.
\item The $1^3D_2$ state of bottomonium has been identified in a 4--step cascade with $M=10161.1\pm0.6\pm1.6$ MeV \cite{y1d}.
\item $\Upsilon(1S)$ decays to $X+(J/\psi,\psi(2S),\chi_{c1},\chi_{c2})$ have been measured \cite{y1scc}. The measured branching fraction is $B(\Upsilon(1S)\to J/\psi + X)=(6.4\pm0.4\pm0.6)\times10^{-4}$.  The measured branching fraction ratios to $B(\Upsilon(1S)\to J/\psi + X)$ for $\Upsilon(1S)\to(\psi(2S),\chi_{c1},\chi_{c2}) + X$ are $0.41\pm0.11\pm0.08$, $0.35\pm0.08\pm0.06$, and $0.52\pm0.12\pm0.09$, respectively.
\item Precision measurements of $\Upsilon(1S,2S,3S)\to\mu^+\mu^-$ have been made, with the result that $\mathcal{B}(\Upsilon(2S)\to\mu^+\mu^-)$ and $\mathcal{B}(\Upsilon(3S)\to\mu^+\mu^-)$ are 56\% and 32\% larger, respectively, than their current PDG values \cite{ymumu}.
\item The radiative decays of bottomonium 1P and 2P states have been measured with improved precision \cite{chibrad}.  While the branching ratios for $\chi_{bJ}(1P)$ states are found to be in good agreement with the current PDG values, those for $\chi_{bJ}(2P)$ are found to be $\sim30-40\%$ larger.
\end{itemize}
Postscript for bottomonium:  No new data taking at the bound bottomonium resonances is expected at any $e^+e^-$ collider, although much remains to be explored.  As an example, none of the spin singlet states of bottomonium, not even the ground state $\eta_b(1S)$, have been identified so far.






\bibliographystyle{aipproc}   


\end{document}